\documentstyle[11pt,aaspp]{article}
\input psfig.tex

\newcommand\qvec{{\bf q}}

\newbox\grsign \setbox\grsign=\hbox{$>$} \newdimen\grdimen \grdimen=\ht\grsign
\newbox\simlessbox \newbox\simgreatbox
\setbox\simgreatbox=\hbox{\raise.5ex\hbox{$>$}\llap
     {\lower.5ex\hbox{$\sim$}}}\ht1=\grdimen\dp1=0pt
\setbox\simlessbox=\hbox{\raise.5ex\hbox{$<$}\llap
     {\lower.5ex\hbox{$\sim$}}}\ht2=\grdimen\dp2=0pt

\begin{document}

\hskip 4in OSU-TA-28/96 

\title{TESTING TREE-LEVEL PERTURBATION THEORY FOR LARGE-SCALE
STRUCTURE WITH THE LOCAL LAGRANGIAN APPROXIMATION}

\author{Zacharias A.M. Protogeros}
\affil{Department of Physics, Ohio State University, Columbus, OH 43210}

\author{Adrian L. Melott}
\affil{Department of Physics and Astronomy, University of
Kansas, Lawrence, KS 66045}

\author{Robert J. Scherrer}
\affil{Department of Physics and Department of Astronomy, Ohio State University, Columbus, OH 43210}

\affil{E-mail: zack@mps.ohio-state.edu, melott@kusmos.phsx.ukans.edu, scherrer@mps.ohio-state.edu}

\begin{abstract}
We test tree-level perturbation theory for Gaussian initial conditions
with power spectra $P(k)\propto k^n$ by comparing the probability
distribution function (PDF) for the density
predicted by the Local Lagrangian Approximation (LLA) with the results
of numerical gravitational clustering simulations.
Our results 
indicate that our approximation correctly reproduces the evolved density PDF 
for $-3 \leq n \leq-1$ power spectra up to the weakly nonlinear regime, while it 
shows marginal agreement for power indices $n=0$ and $+1$ in the linear 
regime and poor agreement beyond this point.
This suggests that tree-level 
perturbation theory (as realized in the Local Lagrangian Approximation)
can accurately predict the 
density distribution function for $-3 \leq n \leq -1$ but fails for $n \ge 0$.
\end{abstract}

\keywords{galaxies: clustering, large-scale structure of universe}

\section{Introduction}
One of the focal points in the study of large scale structure has been the 
evolution of $P(\rho)$,
the one-point probability distribution (PDF) of the density field
(Kofman et al. 1994, Juszkiewicz et al. 1994, Bernardeau and Kofman 1995, 
Protogeros and Scherrer 1996). Whereas in the linear regime, assuming 
Gaussian initial conditions, $P(\rho)$ scales up self-similarly by $D(t)$, 
the growing mode solution, this is not the case once entering the weakly 
non-linear regime, defined by 
density contrast $\sigma=\left<\delta^2={\left( {{\rho-\bar\rho}\over 
{\bar\rho}} \right)} ^2 \right>^{1/2}\approx 1$. 
This behavior may be attributed to coupling of 
different Fourier modes, which in the linear regime evolved independently,
due to effects of non-local interactions in the density field evolution.
Furthermore, multistreaming is expected to become important at this stage,
contributing to the non-linear evolution of the PDF.

Major progress has been made in the past few years on understanding
the quasi-linear evolution of the PDF, i.e., the evolution for $\sigma \la 1$.
In particular, a formalism has been developed by Bernardeau (1992)
to calculate, to lowest order, the hierarchical amplitudes of the
evolved density field (see the next section for a discussion).
More recently,
Protogeros and Scherrer (1996, PS hereafter) have derived
an approximate method (the Local Lagrangian Approximation or LLA), which
provides a simple analytic expression for the evolved PDF which
reproduces, nearly exactly, the tree-level hierarchical amplitudes
of the ``true" evolution.  In PS, the predictions of the Local
Lagrangian Approximation were compared with the results of
the ``exact" gravitational evolution as calculated numerical in a gravitational
clustering code.  For Gaussian initial conditions with tophat smoothing
and a scale free $n=-1$ initial power spectrum, the LLA predictions
were found to be in excellent agreement with the ``true" evolution.
In this paper, we extend this comparison to a range of power spectra:
$-3 \le n \le +1$.

Our motivation for undertaking this study is two-fold:
i)  On a practical level, the Local Lagrangian Approximation appears
to provide an amazingly simple description of the evolution of the PDF
for Gaussian initial conditions.  We wish to determine if this method
is accurate for all Gaussian initial conditions, or only for certain power
spectra.
ii) From a theoretical point of view, the accuracy of the Local Lagrangian
Approximation can also be considered a test of tree-level perturbation
theory, since the LLA reproduces, nearly exactly, all of the hierarchical
amplitudes of the ``true" final density field.  Previous studies
(references) have examined the validity of tree-level perturbation theory
and the importance of higher-order contributions for a few low-order
cumulants of the final density field, such as the variance and the skewness.
However, the LLA provides a method to test {\it all} of the hierarchical
amplitudes at once.

We present a short review of the LLA scheme in Section 2
and a description of the N-body simulations and our comparative results 
in Section 3. Our conclusions are presented in Section 4.  We find
that the Local Lagrangian Approximation (and, therefore, tree-level perturbation
theory)
accurately predicts the evolution
of the density PDF for initial power spectra with $n \le -1$, but
fails for $n \ge 0$.

\section{The Local Lagrangian Approximation}

The Local Lagrangian Approximation is based on the idea that the density
at a Lagrangian point $\qvec$ at a time $t$ can be approximated as a function
only of $t$ and the initial value of $\rho(\qvec)$:
\begin{equation}
\label{localdef}
\rho(\qvec,t) = f(\rho(\qvec,t_0),t).
\end{equation}
If we define $\eta = \rho /\bar \rho$, where $\bar \rho$ is the mean
density, then we choose a mapping of the form:
\begin{equation}
\label{local}
\eta(\qvec,t)={{\eta_0(\qvec)}\over {[1-D(t)\delta_0(\qvec)/\alpha]^{\alpha}}}
\label {eta}
\end{equation}
The mapping in equation (\ref{local}) is related to the Zeldovich
approximation in the sense that $\alpha = 1$ corresponds to the
limit of planar collapse (in which the Zeldovich approximation
gives an exact description of the evolution of the PDF) and
$\alpha = 3$ corresponds to the spherical collapse in the Zeldovich
approximation.

In PS, it was shown that the generating function for the hierarchical
amplitudes is related in a trivial way to the mapping given in equation
(\ref{localdef}).  In particular, if we take $\alpha = 3/2$, the
hierarchical amplitudes for the final density field produced by
the mapping in equation (\ref{local}) will be almost exactly
equal to the true amplitudes produced by exact evolution
(see PS and earlier work by Bernardeau 1992 and Bernardeau \& Kofman 1995).
The resulting PDF, $P(\eta)$, which is derived by applying equation
(\ref{local}) to an initially Gaussian density distribution, automatically
satisfies the normalization condition $\int P(\eta) \eta d\eta = 1$,
but it fails to satisfy $\int P(\eta) d \eta = 1$; this is related,
at some level, to the problem of multistreaming (see PS for the details).
To correct this problem, we multiply equation (\ref{local}) by
a time-dependent normalization factor $N(t)$:

\begin{equation}
N(t)=\left< {1\over {\eta(\qvec,t)}} \right>,
\label{norm1}
\end{equation}    
leading to the LLA mapping:
\begin{equation}
\eta(\qvec,t)= {{{ \left<{ \left( 1-2\delta_l(\qvec,t)/3 \right)} \right>}^
{3/2}}\over {{\left|1-2\delta_l(\qvec,t)/3\right|}^{3/2}}},
\label{LLA}
\end{equation}
where $\delta_l$, the linear-evolved initial density fluctuation,
is simply $\delta_l = D(t) \delta_0$.

This expression gives the final density for an unsmoothed density
field, and so cannot be compared directly with either observations
or gravitational clustering simulations.  If we smooth
the final density field with a spherical tophat window function,
then we obtain a new ``smoothed" local Lagrangian mapping, $f_s$,
given by (Bernardeau 1994; PS):
\begin{equation}
\label{smooth}
\eta_s=f_s(\delta_l)=f\left[ \delta_lf_s(\delta_l)^{-{{(n+3)}\over 6} }\right],
\end{equation}
where we have assumed a power-law power spectrum $P(k) \propto k^n$, and
the smoothed mapping $f_s$ must then be multiplied by the normalization
factor given in equation (\ref{norm1}).  Note that the mapping given in equation
(\ref{smooth}) is not the smoothed density field; rather, it is a density
field which is guaranteed to give the same hierarchical amplitudes
as the ``true" evolved density field.

Combining equations (\ref{LLA}) and (\ref{smooth}), we obtain, for Gaussian initial
conditions with spherical tophat smoothing, the final PDF:
\begin{equation}
P(\eta)d\eta={ 1 \over {N^2}} g({\eta\over {N}}) d\eta 
\label{psmooth}
\end{equation}
where the function $g$ is given by
\begin{eqnarray}
g(x) & = & {\alpha\over {\sqrt{2\pi}\sigma_l}}\left[e^{{-{\alpha^2\over 
{2\sigma_l^2}}} {\left( x^\beta-x^{ {-{1\over{\alpha}}}+\beta}\right)}^2  }
\left[\beta x^{\beta-2}-\left(\beta -{1\over {\alpha}} \right)
x^{-{1\over {\alpha}}+\beta-2} \right]\right. \nonumber\\
 & + & \left. e^{{-\alpha^2\over {2\sigma_l^2}} {\left( x^\beta+
x^{ {-{1\over{\alpha}}}+\beta}\right)}^2  }\left[\beta x^{\beta-2}+
\left(\beta -{1\over {\alpha}} \right)x^{-{1\over {\alpha}}+\beta-2} \right] 
\right].
\label{normexp}
\end{eqnarray}
Here $\alpha = 3/2$, $\sigma_l$ is the linear-evolved rms fluctuation:
$\sigma_l = D(t) \sigma_0$,
and we have defined
\begin{equation}
\beta \equiv {{n+3}\over 6}
\end{equation}
The normalization factor $N$ is given by:
\begin{equation}
N=\int_0^\infty g(x)dx.
\label{norm}
\end{equation}

Note that the second term in equation (\ref{normexp}) corresponds
to the case where the argument of the absolute value in equation
(\ref{LLA}) is negative; this term is negligible compared with
the first term as long as $\sigma_l \la 1$.  Dropping
this second term, we can express the PDF in our Local Lagrangian
Approximation in the particularly
simple form:
\begin{equation}
P(\eta)d\eta={1\over \eta}{1\over {\sqrt {2\pi}\sigma_l}}e^{-{\delta_l^2(\eta)
\over{2\sigma_l^2}}}d\delta_l(\eta),
\label{finp}
\end{equation}
where $\delta_l(\eta)$ is given by
\begin{equation}
\delta_l={3\over 2}\left[ \left(\eta\over N\right)^\beta-
\left(\eta\over N\right)^{-2/3+\beta} \right],
\label{delta}
\end{equation} 
and the normalization factor $N$ is:
\begin{equation}
N={1\over {\sqrt {2\pi}\sigma_l}}\int_0^\infty ({N \over \eta})
e^{-{\delta_l^2(\eta)
\over{2\sigma_l^2}}}d\delta_l(\eta).
\label{normf}
\end{equation}
Note that equation (\ref{normf}) can be integrated by making the change
of variables $x = \eta /N$, which eliminates $N$ from the integral.

\section{N-body Simulations and Results}

We calculate the time evolved density PDF in 
collisionless (Melott et al. 1996) gravitational
clustering simulations of a pressureless dust, $\Omega=1$ universe
described more fully elsewhere (Melott and Shandarin 1993).
We use
$128^3$ particles on a $128^3$ grid and assign initial spectra to have
the form $P(k)\equiv \left<\left| \delta_k \right|^2\right>\propto k^n$, 
with $n=-3,-2,-1,0,+1$ respectively. The initial
spectrum high frequency cutoff is given by the Nyquist frequency of the 
simulation cube $k_{Ny}=64k_f$, where $k_f=2\pi/L$ is the fundamental mode of
the cube and the simulations are terminated at a scale of nonlinearity 
$k_{nl}=16k_f$. The initial density contrast variance $\sigma_0$ is calculated
in tophat smoothed spheres of radii $\lambda=1,2,4,8,16$ cells and the linear
variance corresponding to the final field is obtained from $\sigma_{l}=
(a_{nl}/a_0)\sigma_0$, where $a_{nl}/a_0$ is the ratio of the expansion 
factors at the end and at the beginning of the simulation respectively. We 
use a cloud-in-cell binning for the final densities 
and consider only $\eta$ in the range $[0...4]$. The simulation PDF $P_s(\eta)$ is 
defined as the fraction of the number of volumes of a specific $\Delta\eta$ range, 
$N_{B}$, over the total number of grid volumes $N_{T}$ in the simulation,

\begin{equation}
P_s(\eta)= {N_{B}\over {N_{T}\Delta\eta}}.
\label{psim}
\end{equation}

We expect the rms fluctuation
associated with each measured $P_s(\eta)$ value to scale as the
square root of $N_B$, provided we take into account only the 
independent number of volumes arising after smoothing, 
$N_{I}=3L^3/{4\pi\lambda^3}$, where the $L$ is the side of the 
simulation box and $\lambda$ is the radius of the tophat smoothing sphere.
One can then express the rms fluctuation in $P_s(\eta)$ as:
\begin{equation}
\sigma_{P_s(\eta)}= {{\sigma_{N_{B}}}\over {N_{I}\Delta\eta}},
\label{sigmap}
\end{equation}
where $\sigma_{N_{B}}=\sqrt {N_{I}P_s(\eta)\Delta\eta}$.
To avoid discreteness effects we only use results corresponding to smoothing 
lengths $\lambda \geq 2$ in our comparisons, keeping in mind though that the 
longer smoothing lengths lead to a smaller number of 
independent volumes within the simulation cube and therefore to larger error 
bars in the evaluation of $P_s(\eta)$. 
We examine the evolved PDF results at $\sigma_l$ values selected in the range  
$0.1<\sigma_l<2.0$ so as to extend our investigation from the linear 
($\sigma_l<0.5$), through the weakly nonlinear ($\sigma_l<1.0$), to the nonlinear 
($\sigma_l>1.0$) regimes.

Our results are presented in Figure 1, for negative power indices, and Figure 2
for $n=0,+1$. The solid lines correspond to the Local Lagrangian prediction, while
the results of the gravitational clustering simulations are given as points with $1-\sigma$ 
error bars. In each graph we quote the
extrapolated linear variance $\sigma_l$ and the corresponding smoothing
scale $\lambda$. 
For the $n=-3,-2$ models we use a smoothing length 
$4\leq \lambda \leq 16$ whereas for the $n=-1,0,+1$ 
spectral indices we use $8\leq \lambda \leq 2$ in order to examine the PDF 
behavior in approximately the same $\sigma_l$ range for both the LLA 
predictions and the N-body simulation results.    
For all the negative power models, as easily seen from our results in Figure 1,
the agreement between the Local Lagrangian predicted $P(\eta)$ and the N-body simulation
$P_s(\eta)$ is reasonably good in the range $0.35<\sigma_l<0.72$ and it starts
breaking down beyond that point, the Local Lagrangian $P(\eta)$ systematically 
overestimating the peak of $P_s(\eta)$ and underestimating the tail.
However, the picture is quite different for the $n=0$ and $n=+1$ power 
spectra. 
Here we have agreement between $P(\eta)$ and 
$P_s(\eta)$ only at very low $\sigma_l$ values. Our results indicate not 
only that the Local Lagrangian Approximation breaks down very early in the linear regime
for these values of $n$, but 
also that its range of validity decreases as the power index moves toward
more positive values.   
%One possible explanation for this behavior could be that our assumption about 
%the smallness of the perturbative expansion parameter, assumed to be 
%$\delta<1$, is not valid for the clustering occurring at these 
%models even for low $\sigma_l$ values due to coupling of the long $k$ modes.

In order to examine the behavior of the higher moments of the
density field,
we calculated the quantity 
\begin{equation}
{{<\delta^2_{s}>-{\sigma_l^2}}\over {\sigma_l^4}},
\end{equation} 
where the nonlinear $< \delta^2_{s} >$ was obtained from our simulations.
We observed a decrease with increasing $n$, a result in agreement with Lokas 
et al. (1996) and explained as the effect of previrialization. Furthermore, 
our N-body simulations indicate that a similar behavior is also exhibited by 
the skewness-related quantity 

\begin{equation}
{{<\delta^3_{s}>-{34\over 7}\sigma_l^4}\over {\sigma_l^6}}.
\label{lokas3}
\end{equation} 

\section{Conclusions}
Despite its simple form, the Local Lagrangian Approximation yields the right predictions
for the evolved $P(\eta)$ in the weakly nonlinear regime ($\sigma_l<0.72$),
but only for the cases where $n \le -1$.  For $n > 0$, the LLA fails to agree
with the simulations except at very early stages in the evolution.
Our results are in rough agreement with those of
Scoccimarro \& Frieman (1996), who examined the contribution
of next-to-leading order terms in the perturbative expansion of
$\langle \sigma^2 \rangle$.  They found that these terms diverge for
$n \ge -1$ and converge for $n < -1$.  Our results support the conclusion that
tree-level perturbation
theory fails for $n \ge 0$.  However, we find that
tree-level theory, as expressed in the LLA, can be applied in the case $n=-1$.
This is not totally
contradicted by the results of Scoccimarro \& Frieman (1996), because
in this case the divergence of the next-to-leading order term in their calculations
is only logarithmic.

To the degree that $P(\eta)$ carries all the information about its higher order moments, 
one would expect the predictive power of the LLA to extend to the calculation of
such moments in its range of validity. Calculation of the skewness based on the LLA 
and including one loop corrections (Scoccimarro 1996)
leads to reasonable
agreement as expected. However, the smallness-of-$\delta$ condition is probably not 
satisfied for the power spectra $n=0,+1$ we examined. Possible reasons for the breakdown
of the perturbative approach may be the strong coupling of the long $k$ modes 
due to non-local interactions, as well as multistreaming which may alter
the picture even at the weakly nonlinear stage (Bharadwaj 1996).

>From a practical point of view, the Local Lagrangian Approximation seems
to provide an excellent prediction for the evolution of the density PDF
in the range $n \le -1$.  Since the power spectrum in the quasilinear
regime is close to $n=-1$ (Klypin and Melott 1992), the LLA may be useful
in comparing with observations.  In particular, it may be possible to
invert our mapping to go from the evolved (observed) 
PDF backwards to the initial conditions.   Since the LLA provides
better agreement with the evolved PDF than the Zeldovich approximation,
it is also worthwhile to determine whether a modification of the Zeldovich
approximation can be derived which corresponds to the mapping in equation
(\ref{eta}).

\begin{figure}
\centerline{\psfig{figure=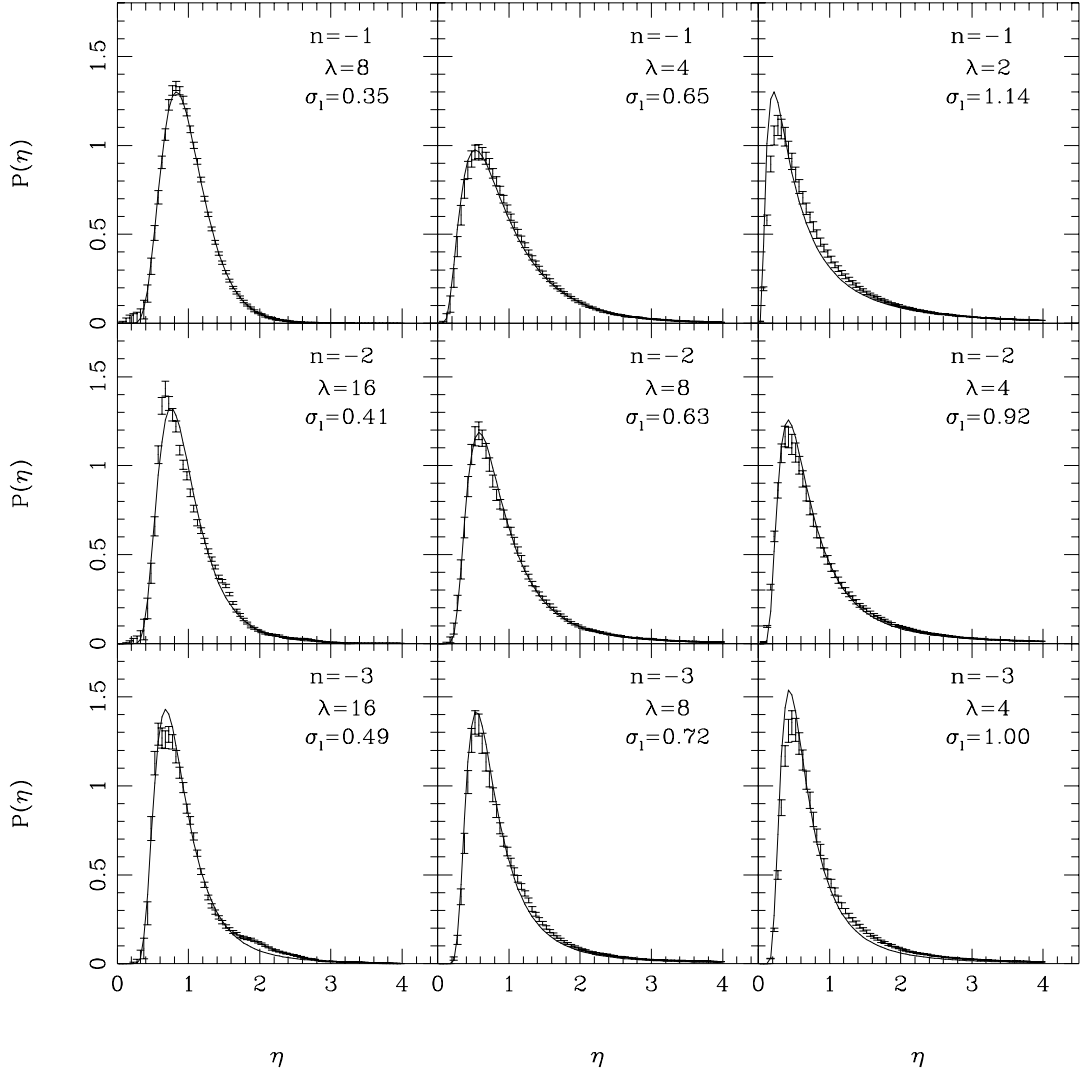,width=6in}}
\caption{\protect\small Comparison of the Local Lagrangian Approximation
top-hat smoothed density 
$P(\eta)$, shown as a continuous line, to the N-body simulation results, shown 
as points with $1-$$\sigma$ error bars, for $\eta={{\rho}/ {\bar\rho}} \le 4$
and $-3 \le n \le -1$.
For each power index $n$ we show three different 
$\sigma_{l(inear)}$ regimes corresponding to smoothing lengths $\lambda$. 
\normalsize}
\end{figure}

\begin{figure}
\centerline{\psfig{figure=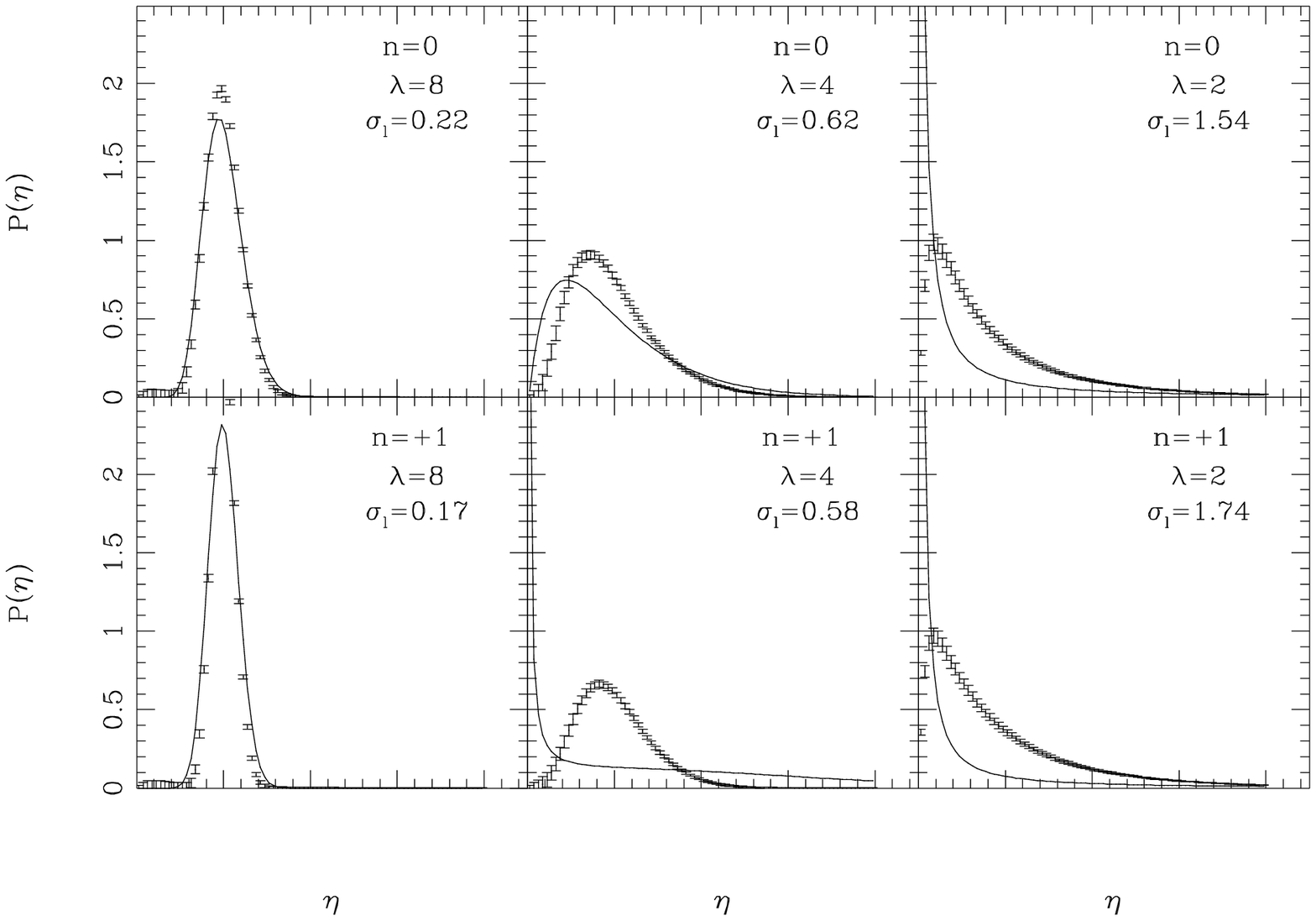,width=6in}}
\caption{\protect\small Same as in Fig.1 but for spectral indices of $n=0,+1$
respectively. 
\normalsize}
\end{figure}

{\bf Acknowledgments}

Z.A.M.P. would like to thank D. H. Weinberg and R. Scoccimarro
for enlightening discussions.
Z.A.M.P. and R.J.S. were supported in part by the
Department of Energy (DE-AC02-76ER01545).  R.J.S was
supported in part
by NASA (NAG 5-2864 and NAG 5-3111).
A.L.M. acknowledges the financial support of
NASA for grant NAGW 3832, the
NSF--EPSCoR program, and the resources of the National Center for
Supercomputing Applications.

\end{document}